\documentclass[10pt]{iopart}
\usepackage{cite}

\newcommand{\mean}[1]{\left\langle #1 \right\rangle}
\newcommand{\smean}[1]{\langle #1 \rangle}

\begin{document}

\title[Azimuthally sensitive correlations]%
      {Characterization and analysis of azimuthally sensitive correlations}

\author{N Borghini}

\address{Service de Physique Th\'eorique, CEA Saclay, 
  F-91191 Gif-sur-Yvette cedex, France}

\begin{abstract}
A unified framework for describing the azimuthal dependence of two-particle 
correlations in heavy-ion collisions is introduced, together with the methods 
for measuring the corresponding observables. 
The generalization to azimuthal correlations between more than two particles is 
presented. 
\end{abstract}

The high amount of data that modern nucleus-nucleus experiments can 
accumulate allow novel types of measurements, giving new information on the 
physics involved. 
An emerging class of experimental analyses thus consists in studying the 
dependence in azimuth of various observables in non-central collisions, 
investigating correlations with the impact-parameter direction. 

A celebrated example of such a correlation with the reaction plane is the 
azimuthal dependence of single-particle emission, the so-called (one-particle) 
anisotropic flow (for a review, see~\cite{Ollitrault:1997vz}).
In the following, we shall focus on the dependence with the azimuth of 
{\em two-particle\/} production. 
If the single-particle production is azimuthally dependent, the two-particle 
production depends on the azimuth as well.
Beyond that, two arbitrary particles may also be correlated, for various 
reasons: 
  a) they may be the decay products of a short-lived particle that decays 
  before reaching the detectors, or
  b) they may both belong to a (di)jet originating from a hard parton 
  scattering, or 
  c) they may be identical particles whose wave-functions interfere. 
Whatever the physical mechanism involved, the resulting two-particle 
correlation will depend on the azimuths of the particles if the short-lived 
parent particles flow (case a), or because the anisotropy of the interacting 
region results in anisotropic patterns of parton energy loss~\cite{Wang:2000fq} 
(case b) and interferometry~\cite{Tomasik:2002rx} (case c). 

The purpose of this paper is to present {\em model-independent\/} observables 
that characterize in a general way azimuthally-sensitive two-particle 
correlations, without any prejudice on the underlying physical mechanism~\cite{%
  Borghini:2004ra}.
We then discuss the experimental measurement of these observables, i.e., the 
measurement of the particle-pair distribution with respect to the reaction 
plane. 
In particular, we shall mention methods of analysis that do not require to 
estimate the reaction plane, in opposition to those currently used~\cite{%
  Adler:2002pb,Heinz:2002au,Adams:2004wz}.
Finally, we comment on the last step of the analysis, namely relating the 
observables we propose to models of the two-particle correlations. 
Quite obviously, that procedure is model-dependent, and we shall present 
several possible ways to tackle the issue. 
We shall conclude by generalizing the overall approach to correlations between 
more than two particles.

\section{Two-particle anisotropic flow}
\label{s:vn-pair}

In this section, we shall introduce general observables for decribing the 
azimuthal dependence of two-particle correlations. 
A convenient approach consists in starting from the azimuthal dependence of 
one-particle emission, i.e., single-particle anisotropic flow; 
the generalization to two-body correlations then follows in a natural way. 

Consider particles of a given type in a given rapidity $y$ and transverse 
momentum $p_T$ window. 
Denoting by $\Phi_R$ the orientation of the reaction plane (throughout the 
paper, azimuthal angles are measured with respect to a fixed direction in the 
laboratory), the probability distribution of the particle azimuth $\phi$ may be 
written as a Fourier series~\cite{Voloshin:1994mz}:
\begin{equation}
\label{p1(phi)}
p(\phi-\Phi_R) = 
\frac{1}{2\pi}\!\sum_{n=-\infty}^{+\infty} v_n \rme^{\rmi n(\phi-\Phi_R)}. 
\end{equation}
The azimuthal dependence of the particle distribution is thus entirely 
characterized by the coefficients $v_n=\smean{\rme^{-\rmi n(\phi-\Phi_R)}}$, 
where angular brackets denote an average over particles and events. 
Note that the normalization choice implies $v_0=1$, while $v_{-n}=(v_n)^*$, 
where $^*$ denotes the complex conjugate, since $p(\phi-\Phi_R)$ is real. 

If the system is symmetric with respect to the reaction plane, 
equation~(\ref{p1(phi)}) reads 
\begin{equation}
\label{p1(phi)bis}
p(\phi-\Phi_R) = 
\frac{1}{2\pi}\left[1+2\!\sum_{n=1}^{+\infty}v_n \cos n(\phi-\Phi_R)\right],
\end{equation}
where $v_n=\smean{\cos n(\phi-\Phi_R)}$ is now a real number. 

It is important to realize that characterizing anisotropic flow by the Fourier 
coefficients $v_n$ is a significant improvement with respect to older 
parameterizations. 
Thus, even though the azimuth of the impact parameter, $\Phi_R$, cannot be 
measured in each event, nevertheless the first $v_n$ can be reconstructed with 
accuracy, by performing a statistical analysis of the multiparticle azimuthal 
correlations between the detected particles (see section~\ref{s:analysis}). 
The distribution $p(\phi-\Phi_R)$, on the other hand, cannot be measured 
accurately---this reflects the fact that for large $n$, the uncertainty on 
the extraction of $v_n$ becomes large~\cite{Ollitrault:1997di}. 
In addition, since $v_n$ is an average quantity, it is easier to compute
theoretically than the probability distribution for a definite azimuth with 
respect to the impact parameter, which varies considerably from event to event 
in computer simulations. 

Keeping these advantages of Fourier coefficients in mind, we can now turn to 
azimuthally-sensitive two-particle correlations. 
This amounts to studying the distribution of {\em particle pairs\/} with 
respect to the reaction plane. 
Now, dropping for the sake of brevity  rapidities and transverse momenta, a 
pair of particles with given types is described by two azimuthal angles 
$\phi_1$ and $\phi_2$.
These can be combined into a ``pair angle''
\begin{equation}
\label{phi_pair}
\phi_{\rm pair} \equiv x\phi_1 + (1-x)\phi_2,
\end{equation}
where $0\leq x<1$, and the relative angle $\Delta\phi\equiv\phi_2-\phi_1$ 
(or any similar combination which does not depend on the orientation of 
the pair with respect to $\Phi_R$). 
Choosing the actual value of $x$ depends on the specific problem considered: 
a natural choice in interferometry and short-lived particle flow studies is
that $\phi_{\rm pair}$ represent the azimuth of the total transverse momentum 
${{\bf p}_T}_1+{{\bf p}_T}_2$, while present studies of azimuthal correlations 
between high-$p_T$ particles rather adopt $x=1$, with the convention that 
particle 1 has the largest transverse momentum in the event (leading particle). 
In any case, the whole particle-pair orientation with respect to the reaction 
plane now affects only one azimuthal angle, namely $\phi_{\rm pair}-\Phi_R$. 

Once this change of angular variables has been performed, parameterizing the 
azimuthal dependence of two-particle correlations becomes straightforward. 
Consider a sample of pairs of particles with given types: the two-particle 
distribution depends on the two particle momenta and rapidities, and on the 
pair angle (with respect to the impact-parameter direction) and relative angle. 
Fixing ${p_T}_1$, ${p_T}_2$, $y_1$, $y_2$, and $\Delta\phi$ (in finite-width 
bins), the distribution is a function of $\phi_{\rm pair}-\Phi_R$ only. 
By analogy with the single-particle anisotropic flow, equation~(\ref{p1(phi)}), 
the probability distribution of the particle-pair azimuthal angle can be 
written as~\cite{Borghini:2004ra}:
\begin{equation}
\label{p2(phi_pair)}
p(\phi_{\rm pair}-\Phi_R) = \frac{1}{2\pi}\!\sum_{n=-\infty}^{+\infty}
v_n^{\rm pair} \rme^{\rmi n(\phi_{\rm pair}-\Phi_R)}.
\end{equation}
The ``pair-flow'' Fourier coefficients are given by 
$v_n^{\rm pair}=\smean{\rme^{-\rmi n(\phi_{\rm pair}-\Phi_R)}}$, where the 
average runs over particle pairs within the phase-space window and $\Delta\phi$ 
range selected; in particular, $v_0^{\rm pair}=1$. 
The real-valuedness of the probability distribution implies 
$v_{-n}^{\rm pair}=(v_n^{\rm pair})^*$, paralleling the similar property of 
the coefficients $v_n$. 

However, unlike $v_n$, the pair-flow coefficient $v_n^{\rm pair}$ is not a real 
number in general. 
This is due to the fact that the system is not symmetric under the change of 
$\phi_{\rm pair}-\Phi_R$ into $-(\phi_{\rm pair}-\Phi_R)$ {\em while keeping 
  $\Delta\phi$ constant\/}. 
It follows that the real form of the Fourier expansion of 
$p(\phi_{\rm pair}-\Phi_R)$ also contains sine terms, which could be omitted in 
equation~(\ref{p1(phi)bis}):
\begin{equation}
\label{p2(phi_pair)bis}
\fl
p(\phi_{\rm pair}-\Phi_R) = 
 \frac{1}{2\pi}\!\left( 1+2\!\sum_{n=1}^{+\infty} \left[
   v_{c,n}^{\rm pair} \cos n(\phi_{\rm pair}-\Phi_R) + 
   v_{s,n}^{\rm pair} \sin n(\phi_{\rm pair}-\Phi_R) \right] \right).
\end{equation}
The real coefficients in expansion~(\ref{p2(phi_pair)bis}) are given by 
$v_{c,n}^{\rm pair}=\smean{\cos n(\phi_{\rm pair}-\Phi_R)}$, analogous to 
$v_n$, and $v_{s,n}^{\rm pair}=\smean{\sin n(\phi_{\rm pair}-\Phi_R)}$, which 
has no equivalent in the case of single-particle flow. 
Quite naturally, the real and complex Fourier coefficients are related by 
$v_n^{\rm pair}=v_{c,n}^{\rm pair}-\rmi v_{s,n}^{\rm pair}$. 
We shall further discuss the sine coefficients $v_{s,n}^{\rm pair}$ later in 
section~\ref{s:summary}.

\section{Analyzing two-particle flow}
\label{s:analysis}

We shall now focus on how to measure the pair-flow Fourier coefficients 
$v_{c,n}^{\rm pair}$, $v_{s,n}^{\rm pair}$. 
As in section~\ref{s:vn-pair}, we first consider the case of the one-particle 
flow coefficients $v_n$, recalling existing methods of analysis. 
The methods for analyzing pair-flow coefficients will then emerge as 
straightforward generalizations. 

Present analyses of anisotropic flow make use of various methods, which fall 
into three main categories. 
On the one hand, there are methods that rely on the determination of an 
estimate of the reaction plane~\cite{Danielewicz:1985hn,Ollitrault:1997di,%
  Poskanzer:1998yz}. 
By contrast, the other methods do not require such a step, but are based on a 
study of the azimuthal correlations between either two~\cite{Wang:1991qh} or 
more-than-two particles~\cite{Borghini:2001vi,Bhalerao:2003xf}. 
Instead of characterizing methods according to whether or not they reconstruct 
the reaction plane, we shall see that another, perhaps more relevant way to 
classify them is to oppose two-particle-based methods to multiparticle ones. 

The mostly used methods are those whose first step is to build event-by-event 
an estimate of the reaction plane, the so-called ``event plane''~\cite{%
  Poskanzer:1998yz}. 
Once this has been done and acceptance corrections have been performed to 
flatten its azimuthal distribution, one correlates the event-plane azimuth with 
that of each outgoing particle, thus obtaining the ``differential flow'' 
$v_n(p_T,y)$. 
This is a most natural idea: to measure distributions with respect to the 
impact-parameter direction, first determine the latter, then correlate outgoing 
particles with this direction. 
However, there are two caveats: first, the event plane does not coincide with 
the real reaction plane; this discrepancy is taken into account in actual 
analyses by correcting for the event-plane dispersion on a statistical 
basis~\cite{Ollitrault:1997di}. 
The second, more serious issue, is the basic assumption of these methods, 
namely that {\em all\/} correlations between the event plane and a given 
particle, i.e., actually, all two-particle azimuthal correlations, are due to 
anisotropic flow only (or at least, that other sources of correlations are 
weak~\cite{Danielewicz:1985hn}). 
This assumption also underlies the computation of the event-plane dispersion 
through the help of ``subevents''~\cite{Ollitrault:1997di}. 
There clearly exist other sources of correlations between particles beyond 
flow: kinematic constraints between decay products, quantum interference 
between identical particles, inter- and intra-jet correlations\ldots and these 
were shown to be of the same order of magnitude as the correlations due to flow 
at ultrarelativistic energies\cite{Borghini:2000cm}, which invalidates the 
assumption. 

The second type of method is less intuitive than those reviewed above, but its 
principle is very simple~\cite{Wang:1991qh}: 
assuming that all two-body azimuthal correlations are due to flow (and symmetry 
with respect to the reaction plane), the average 
$\smean{\cos n(\phi_2-\phi_1)}$ factorizes into 
$\smean{\cos n(\phi_1-\Phi_R)}\smean{\cos n(\phi_2-\Phi_R)}$, where $\phi_1$ 
and $\phi_2$ denote the azimuths of two particles from the same event. 
Averaging first over (pairs of) particles in the whole phase space covered by 
the detectors, one obtains $\smean{\cos n(\phi_2-\phi_1)}=(v_n)^2$, where the 
``integrated flow'' $v_n$ is an average value of the corresponding Fourier 
coefficient. 
One can then restrict particle 2 to a small transverse momentum and rapidity 
window while letting particle 1 in the whole phase space, and average over all 
such possible pairs, which yields the differential flow through 
$\smean{\cos n(\phi_2-\phi_1)}=v_n v_n({p_T}_2,y_2)$, where the first $v_n$ 
denotes the previously determined integrated flow. 
Now, to perform these averages in practice, one builds two-particle correlators 
similar to those used in interferometry studies, dividing a distribution of 
``real pairs'' by a distribution of ``background pairs'' made by mixing 
particles from different events (which automatically removes acceptance 
effects)~\cite{Wang:1991qh}. 
The ratio is then an even function of $\phi_2-\phi_1$ whose Fourier 
coefficients are precisely the two-particle averages 
$\smean{\cos n(\phi_2-\phi_1)}$: by fitting the function, one can extract these 
averages, and thus the flow coefficients. 
Unfortunately, one realizes at once that the same problem of two-particle 
nonflow effects as above also plagues the analysis of flow through
this method.\footnote{Note that the error on the flow estimates due to nonflow 
correlations is the same in both types of methods (using event planes or 
two-particle correlators), so that none of these methods is better than the 
others in that respect.}

Multiparticle methods of flow analysis were devised precisely to remedy the 
issue of nonflow correlations~\cite{Borghini:2001vi,Bhalerao:2003xf}.
The idea of these methods is that when one considers {\em cumulants\/} of the 
correlations between an increasing number (four, six, eight) of particles, the 
relative magnitude of nonflow effects decreases (and, in practice, rapidly 
drops) while the magnitude of the correlations due to anisotropic flow grows, 
because flow is a {\em collective\/} behaviour, which affects all particles. 
One can even think of ``infinite-order cumulants'' that reflect collective 
effects only, isolating flow from other correlations: 
this is the purpose of the application of Lee--Yang zeroes to the analysis of 
flow~\cite{Bhalerao:2003xf}.
By measuring cumulants or Lee--Yang zeroes of a properly chosen function, one 
can thus extract estimates of anisotropic flow that are unbiased by nonflow 
effects, i.e., with a smaller systematic error than two-particle methods. 
The price to pay is an increase in statistical uncertainties, but this increase 
is moderate  in most cases, especially since cuts in phase space so as to 
diminish unwanted correlations are not necessary. 
As the two-particle correlation studies, multiparticle methods proceed in two 
successive steps, to obtain estimates first of integrated flow (using all 
detected particles in each event without any phase space restriction), then of 
differential flow. 

We discussed methods for measuring single-particle anisotropic flow at length, 
for it turns out that only minor modifications of these methods are needed to 
measure the pair-flow coefficients $v_{c,n}^{\rm pair}$, $v_{s,n}^{\rm pair}$. 
More precisely, whatever the method, the first step is exactly the same: the 
determination of the event plane and its resolution in event-plane-based 
methods or the measurement of (single-particle) integrated flow in the methods 
based on two- or multiparticle correlations should be performed in the same way
as in one-particle flow studies, without any change in the procedure. 

The measurement of $v_{c,n}^{\rm pair}$ is then strongly similar to that of 
differential flow: instead of correlating the azimuth $\psi$ of particles in a 
restricted phase-space bin to that of the event plane (event-plane methods) or 
of all other particles in the event (two- and multiparticle methods), one 
simply replaces $\psi$ by the pair angle $\phi_{\rm pair}$.
For instance, in event-plane methods the cosine coefficient is given by
\begin{equation}
v_{c,n}^{\rm pair} = \frac{\mean{\cos n(\phi_{\rm pair}-\Psi_R)}}%
  {\mean{\cos n(\Psi_R-\Phi_R)}},
\end{equation}
where $\Psi_R$ and $\Phi_R$ denote the azimuths of the event plane and of the 
reaction plane, respectively, while the average runs over pairs in the 
(${p_T}_1$, ${p_T}_2$, $y_1$, $y_2$, $\Delta\phi$) bin under study and over 
events.  

Finally, the only significant change, which as we shall see amounts to 
replacing a cos by a sin, concerns the analysis of the sine coefficients 
$v_{s,n}^{\rm pair}$. 
Within event-plane methods, $v_{s,n}^{\rm pair}$ is given by the ratio of the 
average $\smean{\sin n(\phi_{\rm pair}-\Psi_R)}$ over the resolution. 
With the two-particle-correlation method, the difference with single-particle 
flow analyses is that the correlator is no longer an even function of the 
relative angle $\phi-\psi$ (where $\phi$ is any particle while $\psi$ is 
restricted to a ``differential'' bin). 
It is now a non-even function of $\phi-\phi_{\rm pair}$ (where $\phi$ still is 
any particle in the event but those involved in the pair) whose cosine and sine 
Fourier coefficients are respectively
\begin{equation}
\mean{\cos n(\phi-\phi_{\rm pair})}=v_nv_{c,n}^{\rm pair}\quad\mbox{and}\quad
\smean{\sin n(\phi-\phi_{\rm pair})}=v_nv_{s,n}^{\rm pair},
\end{equation} 
where $v_n$ denotes the single-particle integrated flow. 
In the multiparticle method of Lee--Yang zeroes, the necessary modification is 
once again minor: referring the reader to reference~\cite{Bhalerao:2003xf} for 
further detail on the implementation, $v_{s,n}^{\rm pair}$ is given by 
equation~(12) provided one replaces $\cos mn(\psi-\theta)$ in the numerator by 
$\sin mn(\phi_{\rm pair}-\theta)$. 
If one wants to employ multiparticle cumulants, the change is slightly more 
important: to obtain $v_{s,n}^{\rm pair}$, one must consider the imaginary 
parts (replacing $\psi$ by $\phi_{\rm pair}$) of the cumulants defined by the 
power expansion of equations~(26-27) in reference~\cite{Borghini:2001vi}, 
instead of the real parts.
All in all, however, whatever the method used, measuring pair flow does not 
represent a much greater difficulty than measuring single-particle flow, and 
ideally both measurements could be performed at once, since they share the same 
first step. 

Although we emphasize that {\em any\/} method that can be used to measure 
single-particle anisotropic flow $v_n$ can also be applied (modulo minor 
changes) to the measurement of two-particle flow, let us nevertheless make a 
further comment. 
As mentioned above, the various methods relying on two-particle correlations, 
including event-plane methods, assume that all correlations between two 
arbitrary particles in an event are caused by the correlation of each one to 
the reaction plane, i.e., one-particle anisotropic flow. 
If the purpose is to measure {\em single\/}-particle flow, the assumption
introduces systematic errors in the determination of $v_n$. 
However, if the methods are employed so as to derive the {\em pair\/}-flow 
coefficients $v_{c,n}^{\rm pair}$, $v_{s,n}^{\rm pair}$, i.e., azimuthally 
dependent two-particle correlations, there is a logical inconsistency, since 
at some point it is supposed that such effects are negligible. 
Therefore, a logically coherent analysis should rather adopt multiparticle 
methods, despite the larger statistical uncertainty, making use of two-particle 
approaches to estimate systematic errors.

\section{Summary and applications}
\label{s:summary}

In the preceding two sections, we have introduced a set of new observables, the 
pair-flow Fourier coefficients $v_{c,n}^{\rm pair}$ and $v_{s,n}^{\rm pair}$, 
that characterize azimuthally-sensitive two-particle correlations in a unified
way, irrespective of the physics that generate the correlations. 
We then showed how to modify methods of single-particle flow analysis so as to 
measure these coefficients experimentally, again in a model-independent manner. 

To gain some insight on the physics behind these two-particle correlations, a 
further step is needed, which we shall only sketch because it introduces some 
model-dependence which we deliberately want to avoid here. 
For this last step, two general approaches are possible. 
A first possibility is to start from already existing models of correlations, 
which make predictions for definite quantities, and to try to deal with the 
pair-flow observables so as to relate them to these quantities. 
This was attempted in reference~\cite{Borghini:2004ra}, where we showed how to 
use the two-particle flow observables to recover previously-used quantities in 
models of quantum correlations~\cite{Heinz:2002au} or high-$p_T$ jet-like 
correlations~\cite{Adams:2004wz}.
The second approach consists in predicting directly the values of the pair-flow
coefficients $v_{c,n}^{\rm pair}$ and $v_{s,n}^{\rm pair}$ within the framework 
of given models, relating their behaviour to the various parameters of the 
models. 
An instance of such a prediction can be found in reference~\cite{%
  Borghini:2004ra}, where an identity relating together the various 
coefficients was derived in the case of correlations at large momentum between 
(jet) particles originating from high-energy partons that suffered in-medium 
energy loss. 

Other similar predictions can be made on quite general grounds like symmetry 
properties. 
In particular, since the sine coefficients $v_{s,n}^{\rm pair}$ are a novel 
feature that appears in two-particle anisotropic flow while being absent in 
single-particle flow, they are worth some further discussion in various 
physical situations. 

Consider first the (one-particle) anisotropic flow of short-lived particles 
which rapidly decay into two daughter particles that reach the detector, e.g., 
$\rho\to\pi^+\pi^-$.%
\footnote{This can obviously also apply to particles that live longer, but 
  still decay before they reach the detector, as $\Lambda\to p\pi^-$ or 
  $K^0_S\to\pi^+\pi^-$. 
  For such particles, which are usually identified through their decay 
  topology~\cite{Adler:2002pb}, at the expense of cuts which decrease the 
  available statistics, one should correlate {\em all\/} pairs of possible 
  daughters, without cuts. 
  This results in an increased background (and smaller signal-to-background 
  ratio) and the loss of event-by-event identification of the parent particles; 
  but it also means larger statistics and thus smaller statistical errors on 
  the final anisotropic flow values if the background is properly dealt with.}
The $\rho$ mesons can only be reconstructed on a statistical basis, by 
correlating pion pairs, and extracting the $\rho$ flow can be done only by 
inspecting the azimuthal dependence of two-particle correlations between pions 
(sorted according to the pair invariant mass, rather than using the relative 
angle). 
In this specific case, if $\phi_{\rm pair}$ is defined as the azimuth of the 
total transverse pair momentum, i.e., the parent $\rho$ transverse momentum for 
pairs of daughter particles, then the sine coefficients $v_{s,n}^{\rm pair}$ 
vanish when the colliding system is symmetric with respect to the reaction 
plane (the emission of $\rho$ mesons is then symmetric as well, unless parity 
is violated), while the cosine coefficients $v_{c,n}^{\rm pair}$ coincide with 
the $\rho$-meson flow coefficients $v_n$. 

In the case of short-range correlations between identical bosons due to the 
symmetrization of their pair wave-function, the symmetry between both particles 
implies that the sine coefficients will vanish, $v_{s,n}^{\rm pair}=0$. 
Note, however, that Coulomb or strong-interaction correlations will spoil this 
identity, since they break the symmetry between both particles. 

As a final word, let us briefly discuss the generalization to the dependence in 
azimuth of correlations between more than two particles. 
This may be of interest as a way to measure the anisotropic flow of particles 
that are identified in the detectors through at least three particles, as for 
instance $\Xi\to\Lambda\pi$ or $\Omega\to\Lambda K$ followed by 
$\Lambda\to p\pi^-$~\cite{Castillo:2004jy}, or $\omega\to\pi\pi\pi$. 
To tackle such a problem, the recipe is the same as for azimuthally-sensitive 
two-particle correlations: one should first perform a change of angular 
variables so as to isolate the dependence on the orientation with respect to 
the reaction plane in a single azimuth (a ``triplet-angle'') while the other 
variables are {\em relative\/} angles. 
Next, one only needs to consider particle triplets with fixed values of all 
these relative angles as well as of transverse momenta and rapidities, and 
write the three-particle probability distribution, which is now only a function 
of the triplet-angle, as a Fourier series, similarly to 
equations~(\ref{p1(phi)}) and (\ref{p2(phi_pair)}). 
This defines a set of observables, the Fourier coefficients, {\em a priori\/} 
including both sine and cosine terms as in equation~(\ref{p2(phi_pair)bis}).
Any method for analyzing pair anisotropic flow can then be applied to measure 
these three-particle flow coefficients, without modification (apart from the 
obvious replacement of $\phi_{\rm pair}$ by the triplet angle). 
Finally, there enter model-dependent procedures to relate the observables to 
models.

\ack 

I wish to thank the Hot Quarks 2004 organizers for their kind invitation and 
financial support. 
Stimulating discussions with Kirill Filimonov and Jean-Yves Ollitrault are 
gladfully acknowledged.

\end{document}